\documentclass[12pt]{article}

\newif\iffigs\figstrue

\usepackage{epsfig,latexsym}
\usepackage{amsmath}
\usepackage{color}
\usepackage{verbatim}
\usepackage{mathrsfs}
\usepackage{amssymb}
\usepackage[utf8]{inputenc}
\usepackage[T1]{fontenc}

\DeclareMathAlphabet{\mathpzc}{OT1}{pzc}{m}{it}

 \csname
@addtoreset\endcsname{equation}{section}

\def\gz0{\gamma^{0}}

\def\scs#1{\section{\sc #1}}

\def\beq{\begin{equation}}
\newcommand{\eeq}[1]{\label{#1}\end{equation}}
\def\bea{\begin{eqnarray}}
\newcommand{\eea}[1]{\label{#1}\end{eqnarray}}
\def\ba{\begin{array}}
\def\ea{\end{array}}
\def\bec{\begin{center}}
\def\ec{\end{center}}
\def\ba{\begin{align}}
\def\ena{\end{align}}

\def\12{\frac{1}{2}}

\usepackage{slashed}

\usepackage{float}

\usepackage{booktabs}
\usepackage{caption}

\usepackage[mathscr]{euscript}

\usepackage{color}
\usepackage{graphicx}
\usepackage{epsf}
\usepackage{graphicx,epsfig}
\usepackage{amsmath}

\usepackage{multirow}

\usepackage{amsmath, amsthm, amssymb}

\usepackage{epsfig}
\usepackage{cite}
\usepackage{color,colordvi}

\newcounter{hran}

\makeatletter
\renewcommand\section{\@startsection {section}{1}{\z@}%
                               {-3.5ex \@plus -1ex \@minus -.2ex}%
                               {2.3ex \@plus.2ex}%
                               {\normalfont\large\bfseries}}
\makeatother

\vspace{0.5cm}

\newcommand{\bi}{\begin{itemize}}
\newcommand{\ei}{\end{itemize}}

\begin{document}
\thispagestyle{empty}

\vspace{15pt}

\begin{center}


{\Large\sc Supergravity and its Legacy}\\


\vspace{35pt}
{\sc S.~Ferrara${}^{\; a,b,c}$}\\[15pt]

{${}^a$\sl\small Theoretical Physics Department, CERN\\
CH - 1211 Geneva 23, SWITZERLAND \\ }
\vspace{6pt}

{${}^b$\sl\small INFN - Laboratori Nazionali di Frascati \\
Via Enrich Fermi 40, I-00044 Frascati, ITALY}\vspace{6pt}

{${}^c$\sl\small Department of Physics and Astronomy and Mani L.~Bhaumik Institute for Theoretical Physics, \\ U.C.L.A., Los Angeles CA 90095-1547,
USA}\vspace{6pt}

\vspace{8pt}

\vspace{25pt} {\sc\large Abstract}
\end{center}
\baselineskip=14pt
\noindent A personal recollection of events that preceded the construction of
Supergravity and of some subsequent developments.

\vskip 24pt
{\sl
\noindent \small \begin{center} Ettore Majorana Medal address \vskip 12pt Contribution to the Proceedings of the International School of Subnuclear Physics, 54th Course: THE NEW PHYSICS FRONTIERS IN THE LHC-2 ERA, Erice, 14 June - 23 June 2016  \end{center}}
\vfill

\noindent
\baselineskip=20pt
\setcounter{page}{1}

\pagebreak

\newpage
\scs{The Prelude}
In the early 1970s I was a staff member at the Frascati National Laboratories of CNEN (then the National Nuclear Energy Agency), and with my colleagues Aurelio Grillo and Giorgio Parisi we were investigating, under the leadership of Raoul Gatto (later Professor at the University of Geneva) the consequences of the application of “Conformal Invariance” to Quantum Field Theory (QFT), stimulated by the ongoing experiments at SLAC where an unexpected Bjorken Scaling was observed in inclusive electron-proton cross sections, which was suggesting a larger space-time symmetry in processes dominated by short distance physics.

In parallel with Alexander Polyakov, at the time in the Soviet Union, we formulated in those days Conformally Invariant Operator Product Expansions (OPE) and proposed the “Conformal Bootstrap” as a non-perturbative approach to QFT.

Conformal Invariance, OPEs and the Conformal Bootstrap have become again fashionable subjects in recent times, because of the introduction of efficient new methods to solve the “Bootstrap Equations” (Riccardo Rattazzi, Slava Rychkov, Eric Tonni and Alessandro Vichi), and mostly because of their role in the AdS/CFT correspondence.

The latter, pioneered by Juan Maldacena, Edward Witten, Steve Gubser, Igor Klebanov and Polyakov, can be regarded, to some extent, as one of the great legacies of higher dimensional Supergravity. It can be used to gain information on strongly coupled gauge theories, and affords a variety of applications to Particle Physics, String Theory and Condensed Matter Physics.

In 1973 I was awarded a Fellowship at the CERN Theory Division, which was considered the most prestigious place where a young European physicist could go. Subsequent events revealed that going to Geneva was a truly timely choice, since when I started my Fellowship, in the fall of 1973, Julius Wess and Bruno Zumino had just formulated supergauge invariant (now called supersymmetric) QFT in four dimensions, and then found that milder divergences occur in these theories with respect to standard renormalizable QFT.

The original name was inherited from fermionic strings, where a local (conformal) fermionic symmetry on the world sheet was introduced by Andr\'e Neveu and John H. Schwarz, Pierre Ramond, and Jean-Loup Gervais and Benji Sakita.

During my two-year period as a CERN Fellow I wrote four papers with Zumino, two of which also with Wess and Jean Iliopoulos (at the time a visiting scientist at CERN). With Zumino we formulated four dimensional Supersymmetric Yang Mills Theories (the new name for supergauges introduced by Abdus Salam and John Strathdee), thus opening the way to consider supersymmetric extensions of the Standard Model, the ISM and alike.

The merging of supersymmetry with Yang-Mills gauge symmetry was a non-trivial step, and the potential link with the Standard Model was motivated by the so-called “Hierarchy Problem” and “Naturalness of Scales”, which in these new theories were alleviated thanks to so-called “Non-Renormalization Theorems”.

Many people contributed to these developments. Aside from the previously mentioned collaborators of  Zumino, I can surely mention Kelly Stelle, Peter West, Warren Siegel, Jim Gates, Marc Grisaru and Martin Ro{\v c}ek.

In retrospect, the most influential of my joint papers with Zumino, written during my period as a Fellow, was the one on the ``Supercurrent multiplet'', a multiplet which extended Noether’s theorem to Supersymmetry and whose components include the energy momentum tensor and a vector-spinor fermionic current, as well as a vector, a scalar and a pseudo scalar. The considerations that follow were among the starting points which led Daniel Freedman, Peter van Nieuwenhuizen and myself to formulate Supergravity in four dimensions as the supersymmetric extension of General Relativity (GR).

Since in theories with local (gauge) symmetries Noether currents couple, to first order, to gauge fields, the fact that the stress tensor couples to the metric tensor in GR indicated that the spinor current should have coupled to a spin-3/2 massless gauge field, of the type introduced by William Rarita and Julian Schwinger (RS) in 1941, and promptly used by Shuichi Kusaka as a hypothetical spin-3/2 neutrino.

The structure of the super-current multiplet indicated two facts that were later exploited, namely that a supersymmetric extension of GR should have involved only two particles of helicities (2,3/2), and that the field corresponding to the latter should have been a Rarita-Schwinger field, to be coupled to GR. This seemed in contrast with a superspace formulation of a Super Riemannian Geometry, which was originally pursued by Pran Nath, Richard Arnowitt and Zumino, where many ordinary fields with different spins were needed.

In the fall of 1975, even if I had offers from the US, following a suggestion of Jacques Prentki, at the time CERN-TH Division leader, I accepted a CNRS position at the \'Ecole Normale Sup\'erieure in Paris.

There I had the chance to meet Daniel Freedman, who was also trying to work out a consistent theory for spin 3/2 coupled to GR.

Peter van Nieuwenhuizen was then his colleague at Stony Brook. He was also collaborating with us, since he was investigating the quantum properties of fields of different spins coupled to GR, and the possibility of ultraviolet cancellations similar to those encountered in super-Yang-Mills theories.

In the spring of 1976 the breakthrough was achieved. Using the Noether procedure, Fierz rearrangements and Riemann tensor identities (and a computer at the Brookhaven National Laboratory), we proved that there is a unique Lagrangian, with only spin 2 and 3/2, and a unique set of supersymmetry transformation rules, which defines pure Supergravity in four dimensions, the supersymmetric extension of GR.

General coordinate transformations, when combined with Supersymmetry, promoted the latter to a gauge symmetry as well, since spinors in curved space must necessarily depend on the space time point due to the local Lorentz symmetry.

\scs{The Play}

The construction of Supergravity as the gauge theory of local supersymmetry prompted Stanley Deser and Zumino to formulate this theory as the Einstein-Cartan version of GR minimally coupled to a (spinor valued) one-form field, the RS gravitino (word coined by Sidney Coleman).

In their work, a quartic spin-3/2 coupling with gravitational strength, required in our formulation by local supersymmetry, originated as a torsion contribution to the spin-connection when treated as an independent field. This extends the so-called first-order or Palatini formulation of GR.

A subsequent superspace formulation of Supergravity by Wess and Zumino in 1978, where an appropriate super torsion constraint was introduced to reproduce the flat super geometry of the super-Poincar\'e algebra, revealed that our second-order approach, rather than the Palatini-Einstein-Cartan formulation, is equivalent to the Wess-Zumino curved Superspace. In particular, it reproduces the so called “auxiliary fields”, needed to have an off-shell formulation and thus a tensor calculus as in standard GR, introduced by Peter van Nieuwhenuizen and myself and Stelle and West.

In the months following the Spring of 1976 several important results were obtained, both during the Summer Institute organized by the \'Ecole Normale Sup\'erieure in Paris and during subsequent visits of some of us to Stony Brook and to CERN.

In particular, also in collaborations with Eugene Cremmer, Joel Scherk, Bernard Julia, Ferdinando Gliozzi and Peter Breitenlohner, the first matter couplings were obtained and, with Peter van Nieuwnhuizen, the first extended N=2 Supergravity formulated.

In a fundamental parallel work,  Gliozzi, David Olive and Scherk discovered that the dual spinor model, after a GSO projection, becomes supersymmetric in space-time and not only on the string world-sheet. Moreover, they found that N=4 Supergravity Coupled to N=4 Yang-Mills is the low energy effective theory of the 10D dual spinor model. Meanwhile Scherk, Lars Brink and Schwarz classified super Yang-Mills theories in all dimensions.

In these milestones of Superstring Theory, Supergravity in diverse dimensions started to emerge as a low-energy limit.

The years 1977 and 1978 were the most performing for Supergravity because the discovery of off-shell formulations allowed to write general matter-coupled Lagrangians and to begin a systematic study of the SuperHiggs effect, the mass generation for the gravitino as a consequence of the spontaneous breaking of local Supersymmetry. This includes a joint work with Cremmer, Julia, Scherk, Girardello and van Nieuwenhuizen.

Moreover, the ultraviolet properties of Supergravity theories started to be investigated and pure supergravity was shown to be finite at one and two loops. This is to be contrasted with the later, fundamental work of Marc Goroff and Augusto Sagnotti, where GR was shown to diverge at two loops.  Three-loop counter terms in different forms of Supergravity were however exhibited by different authors, including Renata Kallosh, Grisaru and Deser, Jason Kay and Stelle.

In a pivotal step, in those years examples of all extended supergravities in four dimensions were derived and the appearance of electric magnetic (e.m) duality symmetries was noticed, with duality groups that act in a non-linear way on the scalar fields of the theory.

The N=8 supergravity in 4D was constructed by Cremmer and Julia, and shown to possess a very large duality group, the exceptional group E7(7).

Gauged N=8 Supergravity in 4D was constructed by Bernard de Wit and Hermann Nicolai, and gauged Supergravity in diverse dimensions was studied by many people, including Michael Duff, Murat G\"unaydin, Chris Hull, Nicholas Warner, Paul Townsend, Leonardo Castellani, Riccardo D'Auria and Pietro Fr\'e.
E.m. dualities were later promoted to string U-dualities by Hull and Townsend.

Cremmer, Julia and Scherk constructed Supergravity in the highest possible number of dimensions, eleven, following a previous classification of Super-Poincar\'e algebras by Werner Nahm.

This theory led Witten to formulate, in the Mid 1990s, the remarkable M-theory hypothesis, thus identifying an elusive theory in eleven space-time dimensions via the strong coupling limit of the Type-IIA Superstring. Through webs of dualities, this step allowed to unify the different form of 10D Superstrings, thus also connecting different types of extended objects, called generically branes, which emerge as solitonic solutions from higher dimensional Supergravity.

Duff had been the first, in the 1980s, to advocate the central role of generic extended objects in an overall picture.

D-branes in string theories first appeared in the classic works of Massimo Bianchi, Gianfranco Pradisi and Sagnotti, and of Joseph Polchinski.

The advent of M theory marked the “second string revolution”. The 1980s witnessed extensive investigations of Kaluza-Klein (KK) theories, largely aimed at classifying possible KK consistent reductions (spontaneous compactifications).

General Lagrangians including soft breaking terms of gravitational and/or gauge origin were derived by Cremmer, Luciano Girardello, Antoine van Proeyen and myself, and were applied to supersymmetric models of Particle Physics.

No-scale Supergravity, a theory with a naturally vanishing cosmological constant, was built by Cremmer, Costas Kounnas, Dimitri Nanopoulos and myself, and was readily applied to Physics beyond the Standard Model with John Ellis. Twenty years later these models were linked to string vacua in flux compactifications by Steve Giddings, Shamit Kachru and Polchinski.

All these developments made it possible to derive parameter spaces of masses and couplings for supersymmetric extensions of the Standard Model, which can be explored in experimental searches for new physics, as was done at LEP experiments in the past and is currently done at the LHC collider.

Supersymmetric extensions of the Standard Model were introduced and widely investigated by Pierre Fayet and by Savas Dimopoulos and Howard Georgi, relying heavily the soft-breaking terms proposed by Girardello and Grisaru and derived in Supergravity by Riccardo Barbieri, Carlos Savoy and myself, and independently by Lawrence Hall, Joseph Lykken and Steven Weinberg, and by Arnowitt, Ali Chamseddine and Nath.

``Split Supersymmetry'', a mechanism advocated in order to get large splitting among MSSM super partners while keeping the nice predictions of gauge coupling unification was later introduced by James Wells, Nima Arkani-Hamed, Dimopoulos, Gian Giudice, Andrea Romanino, and was further developed by the same authors.

A new wave of activity in Supergravity was spurred in 1984, when Michael Green and Schwarz ignited the ``first superstring revolution'' with the discovery of their cancellation mechanism (for gauge groups E8xE8 or/and SO(32)), making Heterotic and Type-I superstrings (which can only afford the latter option) and their point limit, 10D (1,0) supergravity, free of potential gauge anomalies and also of the gravitational anomalies previously studied by Luis Alvarez-Gaum\'e and Witten.

The anomaly cancellation mechanism placed restrictions on compactification manifolds, uncovering enticing connections between the supergravity analysis and Algebraic Geometry.

The Calabi-Yau compactifications investigated by Philip Candelas, Gary Horowitz, Andrew Strominger and Witten resulted in a paradigmatic example of N=2 supergravity analysis in terms of Special Geometry and Mirror Symmetry.

All this machinery of compactifications was thus reconsidered, taking into account restrictions coming from the Green-Schwarz mechanism. Important implications exist for compactifications to six dimensions preserving eight supercharges, where some anomaly canceling terms modify the low energy couplings of tensor multiplets to gauge fields.

In the 1990s, Supergravity evolved into a major tool for the study of Black Holes (BH), and in particular of their BPS properties, given their emergence as supersymmetric solitons in four-dimensional N-extended Supergravity.

An interesting interplay between charged black-holes and the moduli-space geometry of the underlying Supergravity emerged through the “Attractor Mechanism”, a property noticed by Kallosh, Strominger and myself when we met at an Aspen Summer Institute, in Colorado, in 1995.

In N=2 Supergravity supersymmetric black holes enjoy a universal formula for their Bekenstein-Hawking Entropy, in terms of the (square modulus) of the central charge computed at its extremum in moduli space (see also work with Gary Gibbons).

The name attractors was coined since these solitons follow trajectories in the moduli space of scalar fields which, independently of the initial conditions, reach the same extremal point, which coincides with the value of field at the BH Horizon. This explains why the moduli dependent ADM mass is a continuous parameter while the entropy is quantized in terms of the BH charges.

These results were then extended to theories with N-extended supersymmetry and in higher dimensions in work including my collaborators and friends Kallosh, Paolo Aschieri, Bianca Cerchiai, Laura Andrianopoli, Stefano Bellucci, Anna Ceresole, Chamseddine, Gianguido Dall'Agata, D'Auria, Fr\'e, Murat G\"unaydin, Maria Lled\'o, Alessio Marrani, Mario Trigiante, Armen Yeranyan and Zumino.

Partial supersymmetry breaking in N=2 supergravity was studied in collaboration with Girardello and Massimo Porrati, with inspiration from the work of Ignatios Antoniadis, Herv\'e Partouche and Tomasz Taylor, and a minimal model with flat potential was explored.

In 1997 Supergravity found itself again at center stage in the work of Juan Maldacena, amusingly entitled ``The Large-N Limit of Superconformal Field Theories and Supergravity'', where the famous AdS/CFT duality was formulated, which was soon followed by important additions by Witten, Gubser, Klebanov and Polyakov. This might be called the “third string revolution”, and would have been impossible without a particular compactification of type IIB Supergravity on AdS5 x S5 derived in 1985 by Hyiung-Jin Kim, Larry Romans and Peter van Nieuwenhuizen.

This completes my personal recollection of many past years devoted to Supergravity. The recent years 2013-2016 witnessed a revival of spontaneous breaking of local Supersymmetry, boosted by applications to Cosmology, and in particular to the inflationary Universe with a de Sitter phase, with an eye to observational results of the Planck Mission.

A prominent role in this work was played by Andrei Linde, a proponent of Inflationary Cosmology, and indeed Daniel Freedman and I collaborated independently with Kallosh and Linde. I also collaborated with a number of colleagues, including Antoniadis, Dall'Agata, Emilian Dudas, Alex Kehagias, Porrati, Sagnotti, Jesse Thaler, Van Proeyen, Timm Wrase and Fabio Zwirner.

Supergravity has also found applications in higher-spin field theories, via the AdS/CFT correspondence. Supersymmetric versions of OPEs and of the Conformal Bootstrap for N-extended superconformal field theories are currently under investigation.
In the quantum domain N=8 Supergravity in four dimensions was proved to be ultraviolet finite at very high loop order, following the pioneering work of Zvi Bern, Lance Dixon and David Kosower. These properties seem to descend from some amazing relations with the rigid N=4 Yang-Mills theory, which possesses exceptional conformal and integrability properties. It is not yet excluded that N=8 Supergravity be a perturbatively finite theory of Quantum Gravity. In any case this search is going to teach us new lessons on the ultraviolet cancellations which can occur in theories including General Relativity.

\vskip 24pt

\noindent{\large \bf Acknowledgements}\\ \noindent
The author was supported in part by the CERN Th Department
and by INFN (IS CSN4-GSS-PI).
\end{document}

\end{thebibliography}
\end{document}